\documentstyle[11pt,newpasp,twoside]{article}
\def\edcomment#1{\iffalse\marginpar{\raggedright\sl#1\/}\else\relax\fi}
\reversemarginpar

\begin{document}
\title{The dark matter content of lensing galaxies at 1.5 $R_e$}
\author{Paul L. Schechter}
\affil{Massachusetts Institute of Technology, 
77 Massachusetts Avenue, Cambridge MA 02139, USA}
\author{Joachim Wambsganss}
\affil{Universit\"at Potsdam, Institut f\"ur Physik, 
Am Neuen Palais 10, 14469 Potsdam, Germany}

\begin{abstract}
Many gravitationally lensed quasars exhibit flux ratio ``anomalies''
that cannot be explained under the hypothesis that the lensing
potential is smooth on scales smaller than one kpc. Micro-lensing by
stars is a natural source of granularity in the lens potential. The
character of the expected fluctuations due to micro-lensing depends
sensitively on the relative surface densities of micro-lenses (stars)
and smoothly distributed (dark) matter. Observations of flux ratios
may therefore be used to infer the ratio of stellar to dark matter
along the line of sight -- typically at impact parameters 1.5 times
the half light radius.  Several recently discovered systems have
anomalies that would seem to be explained by micro-lensing
only by demanding that
70-90\% of the matter along the line of sight be smoothly distributed.
\end{abstract}
\section{The Problem}
Schneider (this volume) discusses the flux ratio anomalies observed in
gravitationally lensed quasars and their interpretation as the
consequence of substructure within the intervening galaxy.  The
problem is illustrated by two quadruple systems: the archetype,
PG1115+080 (Weymann et al. 1980), and one recently discovered,
SDSS0924+0219 (Inada et al. 2003). The image configurations are nearly
identical, so it is no surprise that models based on the image
positions predict nearly identical flux ratios.  But as seen in Figure
1, the observed flux ratios are dramatically different.  In particular
the $A1$ and $A2$ images in PG1115 are much more nearly equal in
brightness than the corresponding pair, $A$ and $D$, in SDSS0924.
\begin{figure}
\vspace{3.0truein}
\includegraphics{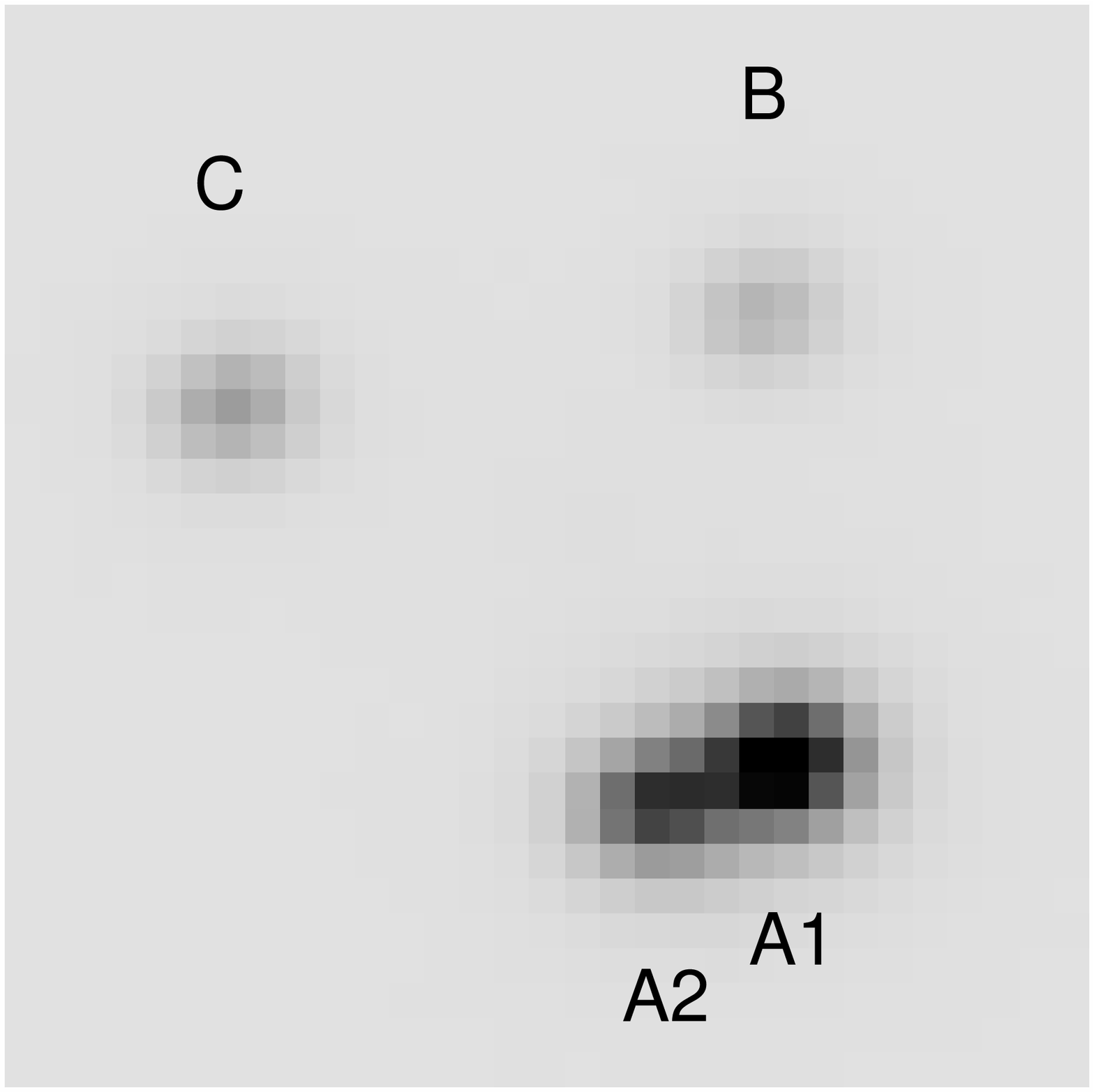}
\includegraphics{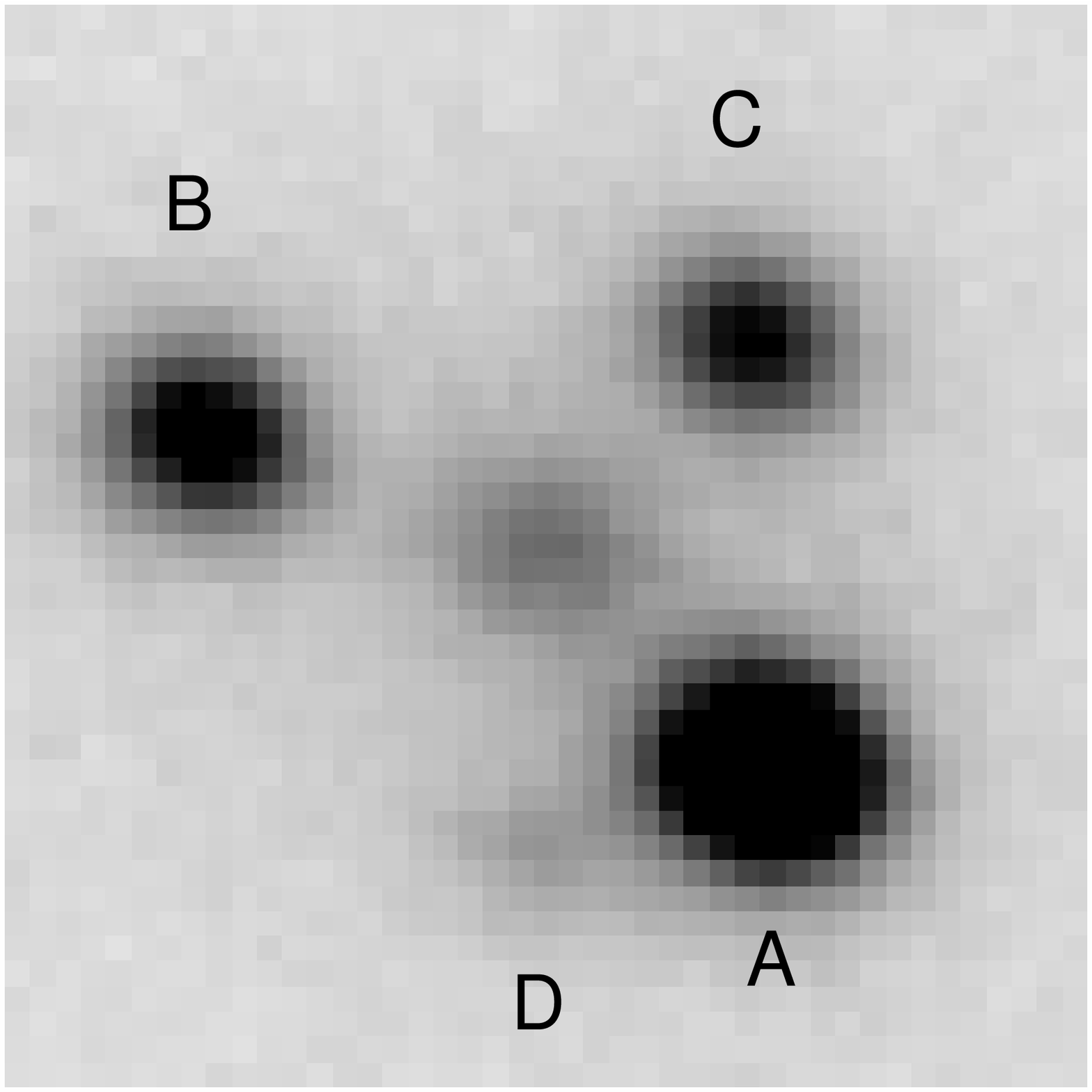}
\caption{Left: An $I$ filter image of PG1115+080 taken with the Baade
6.5-m telescope.  QSO components $A2$ and $A1$ differ by
0.5 mag and are separated by
0\farcs48. Right: A Sloan $i'$ image of SDSS0924+0219, taken with
the Clay 6.5-m telescope, rotated so as to mimic PG1115+080. The
$D$ and $A$ components differ by 2.5 mag and are
separated by 0\farcs66.}
\end{figure}
Metcalf and Zhao (2002) have argued that the two closest images in
PG1115 ought to be yet more nearly equal than they are.  One may
well wonder whether the fluxes in {\sl any} quadruple system have the
expected ratios.

\section{Solution A: Micro-lensing}

Within months of the discovery of the first gravitational lens, Chang
and Refsdal (1979) suggested that the observed flux ratios might be
different from those predicted assuming smooth galaxy potentials owing
to the granularity introduced by individual stars within the galaxy.
There is evidence for such micro-lensing in the temporal behavior of lensed
systems, most spectacularly in the quadruple system Q2237+0305 (e.g.\
Wo\'zniak et al.\ 2002), but also in HE1104-1805 (Schechter et al.\
2003) and, with somewhat less certainty, in another handful of systems
(Wambsganss 2001).  Indeed Vanderriest et al.\ (1986) report that in
the early 1980s, the bright components in PG1115 {\sl were} more
nearly equal than they are in Figure 1, as Metcalf and Zhao's models
would predict.

But while stars might explain observed flux ratio anomalies in a
quasar's optical continuum, they cannot explain anomalies in its radio
continuum.  A micro-lens will cause brightness fluctuations only if
its Einstein ring is larger than the lensed source.  Radio continuum
emission is thought to arise from regions very much larger than the
typical stellar Einstein ring.  The system CLASS 1555+375 (Marlow et
al. 1999) is a radio analog of PG1115, with its two close images
differing by a factor of two in flux.  Something else is needed.

\section{Solution B: Milli-lensing}
Mao and Schneider (1998), Metcalf and Madau (2001), Chiba (2002), and
Dalal and Kochanek (2002) invoke substructure with masses of order
$10^6 M_{\sun}$ to explain radio anomalies.  Such substructure would
have to comprise a few percent of the mass along the line of sight.
While globular clusters and dwarf galaxies occur too infrequently,
N-body simulations (Moore et al.\ 1999; Kravtsov et al.\ 1999)
indicate that galaxies ought to contain dark matter mini-halos capable
of producing such milli-lensing (cf. Wambsganss and Paczy\'nski 1992).
Mao (this volume) discusses at length this interpretation of the radio
anomalies.

\section{Working Hypothesis: Micro-lensing}
\begin{figure}
\vspace{4.0truein}
\includegraphics{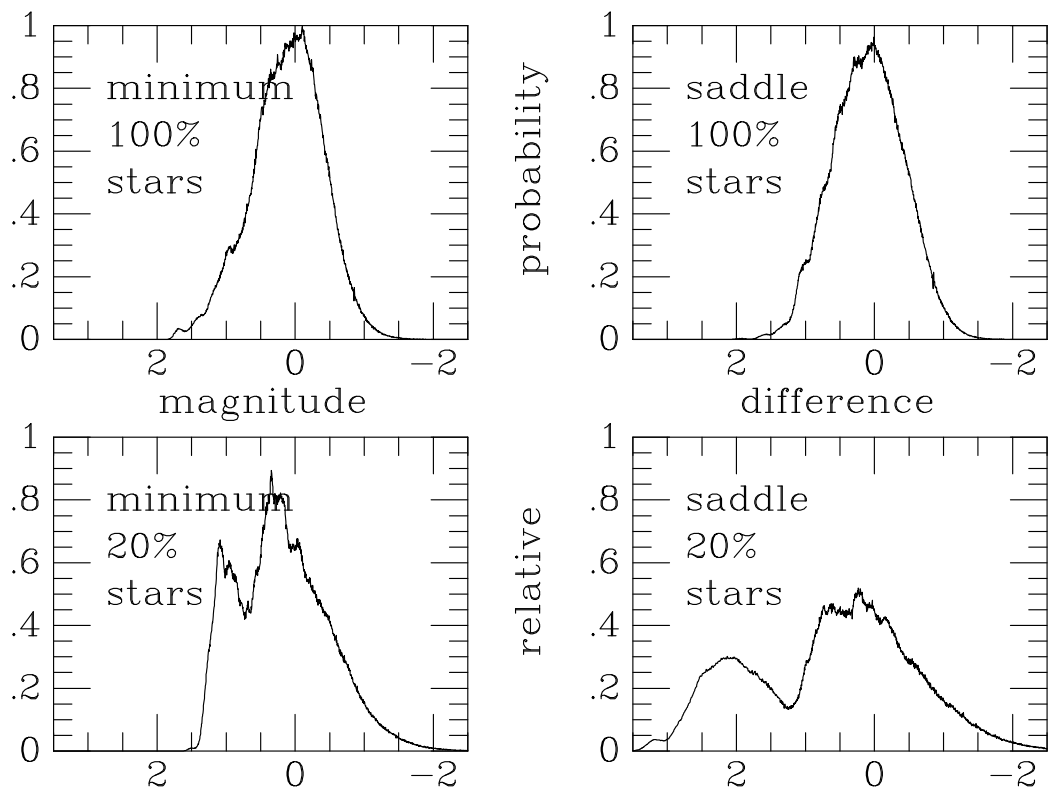}
\caption{Relative magnification probabilities for micro-lensing
simulations appropriate to the close pair of images in SDSS0924+0219
The panels on the left are for the positive parity image, those on the
right for the negative parity image.  The panels on the bottom
assume that 80\% of the surface density is in a smoothly distributed
component, while those on the top take all of the surface density
to be in the form of stars.  The abscissa is zeroed at the predicted
magnification.  Negative abscissas indicate brighter images.
}
\end{figure}

Flux ratio anomalies in the optical continuum may be due either to
micro-lensing or milli-lensing.  One might discriminate between the
two by comparing optical fluxes with radio fluxes, but most quasars are
radio quiet.  Fortunately their are alternatives.  Agol et al.  (1999)
have used mid-IR fluxes, which are also thought to arise from a
relatively large region, instead of radio fluxes.  More recently
Wisotzki et al. (2003) have used broad emission line region (BLR)
fluxes, which again are thought to arise from a much larger region
than the Einstein rings of stars.  For the quadruple system
HE0435-1223, Wisotzki et al. find that the BLR fluxes fit a smooth
model for the system a factor of four better than the continuum
fluxes.  As part of an HST study of just this effect (GO-9854), a
similar result has been found for the system RXJ0911+0551.  Without such
measurements we cannot rule out milli-lensing as the source of the
optical flux ratio anomaly in any given system.  We shall nonetheless
proceed on the working hypothesis that most (though certainly not all)
of what we observe is due to micro-lensing.

\section{Quantitative Micro-lensing}
The positions of the four images of a quadruply lensed system yield
macro-models (e.g. Keeton 2001) for the potential of the lensing
galaxy.  At each image position, the macro-models predict a well
constrained dimensionless mass surface density, $\kappa$, and a shear,
$\gamma$, that measures the differential stretching of the image.  A
predicted magnification for each image is straightforwardly computed
from $\kappa$ and $\gamma$.  The largest source of error in the
magnifications arises from the uncertainties in the radial profile of
the lensing galaxy.

The lens' surface mass density may not be perfectly smooth -- it can
be clumpy on scales small compared to the separation between images.
If the surface density is clumpy on scales greater than or equal to
the size of the emitting region of the quasar, the local values of
$\kappa$ and $\gamma$ in the vicinity of the image (and therefore
governing its actual magnification) will differ from those computed
from the macro-model.

Consistent with our working hypothesis, we assume that a fraction of
the surface density is in stars and that the remainder is in a smooth,
presumably non-baryonic component.  The stellar component is taken to
be randomly distributed, and its statistical effects upon the
magnification of images can be computed using a variety of techniques
(e.g. Paczy\'nski 1986; Wambsganss 1992; Witt 1993).

In Figure 2 we show results from simulations of a close pair of images
with model parameters appropriate to the anomalous pair in SDSS0924.
The smooth predictions would have the two images equally bright.  The
top two panels show the expected magnitude distribution assuming that
100\% of surface density is in stars.  The bottom two panels show the
expected distributions for a stellar fraction of 20\%.  In all cases
we have assumed a point source.

While the two images have rather similar probability distributions
assuming 100\% stars, they are very different assuming 20\% stars.
What distinguishes the two is the parity of the images.  The image
on the left, formed at a minimum of the light time, has positive parity.
The image on the right, a saddlepoint of the travel time, has negative
parity.  Crude explanations of the different behaviors are given by
Schechter and Wambsganss (2002) and by Granot, Schechter and
Wambsganss (2003).

The probability of a given magnitude difference is obtained by
shifting the panel on the left with respect to the panel on the right,
taking the product of the two probabilities, and integrating over all
magnitudes.  A difference as large as 2.5 magnitudes is much more
likely for the bottom two panels than for the top two, but only if the
saddlepoint is fainter than the minimum (as it is).

\section{Measuring the Stellar Fraction}
The different behaviors of the images in a quadruple system as one
varies the stellar fraction suggest a scheme for measuring the stellar
mass fraction along the line of sight.  Measure the fluxes, wait for
the source to move or the stars to reconfigure, measure again, and
accumulate magnification statistics for the four images.  These may
then be compared with predicted histograms for the four images under
various assumptions about the stellar fraction.  Kochanek (2003) has
done something along these lines for the system Q2237+0305.  But this
system is unique in having a short timescale for change; for most
quadruples the timescale for significant change is of order ten years.

\begin{figure}
\vspace{2.875 truein}
\includegraphics{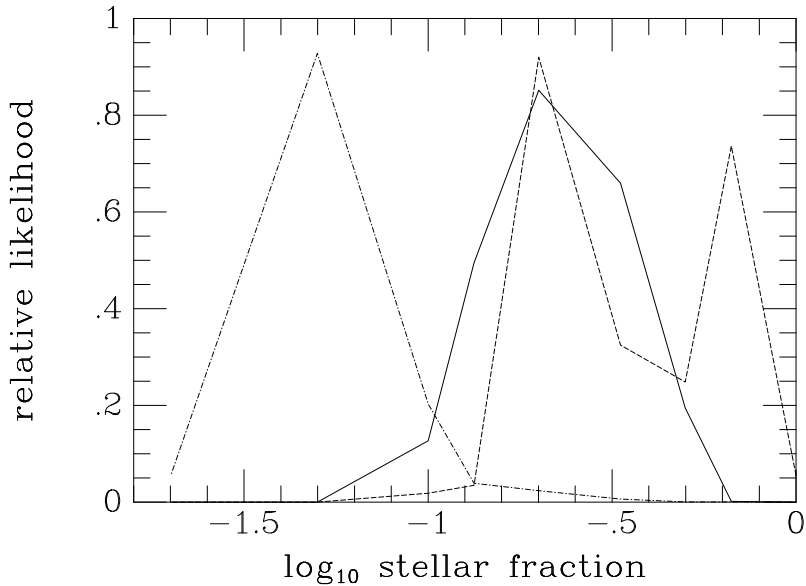}
\caption{Relative likelihood of the stellar mass fraction for
the sample of lenses considered here.  The dot-dashed line
gives the result for all eleven systems assuming that
the lensed QSO is unresolved by the Einstein rings of
the intervening stars.  The solid line assumes partial resolution
of the  QSOs.  The dashed line shows the effect of
eliminating SDSS0924+0219 from the sample.
}
\end{figure}

Alternatively, one can obtain single epoch snapshots of a large number
of systems.  For each system, one calculates (by extension of the
method described above) the probability of the observed flux ratios
assuming some specific stellar fraction.  One identifies the product
of those probabilities with the likelihood of that stellar fraction.

We have carried out this second test for a sample of eleven quadruple
systems with well measured optical continuum fluxes.  The results are
presented in Figure 3.  The dot-dashed line shows the result of
straightforward application of the method, with the likelihood peaking
at a stellar fraction of 5\%.  This seems unlikely, since the stellar
fraction at the position of the images, calculated using the observed
surface brightness profiles of the lensing galaxies and assuming a
``reasonable'' mass-to-light ratio would put 20\% of the mass in
stars.

But we took our quasars to be very nearly point sources.  Had we taken
the angular size of the quasar's continuum emitting region to be
comparable to the Einstein ring of an intervening star, we would
expect smaller fluctuations.  As the fluctuations are largest toward
the middle of the range of stellar fractions plotted (Schechter and
Wambsganss 2002), this drives the peak toward the center of the
figure.  We have plotted results for a very simple extended model, taking
50\% of the optical flux to come from a pointlike source and 50\% from
a source very much larger than the stellar Einstein rings.  The
results, plotted as a solid line in Figure 3, are more in line with
our expectations, and would seem to exclude the possibility that all
of the surface density is in the form of stars.

The likelihoods calculated for individual objects are quite broad.  As
seen from the dashed line in Figure 3, it is largely because of
SDSS0924 that a 100\% stellar fraction is ruled out.  If the large
flux ratio anomaly observed in its optical continuum is {\it not} seen
the emission line flux ratios, this system, by itself, presents strong
evidence for a smooth dark matter component.

\end{document}